\documentclass[fleqn,usenatbib]{mnras}

\usepackage{newtxtext,newtxmath}

\usepackage[T1]{fontenc}

\DeclareRobustCommand{\VAN}[3]{#2}
\let\VANthebibliography\thebibliography
\def\thebibliography{\DeclareRobustCommand{\VAN}[3]{##3}\VANthebibliography}

\usepackage{graphicx}	
\usepackage{amsmath}	

\usepackage{{booktabs}}
\usepackage{graphicx}
\usepackage{fixmath}
\usepackage{soul,xcolor}
\setstcolor{red}
\usepackage{array} 
\usepackage{changepage}
\usepackage{rotating}
\usepackage{comment}
\usepackage{multirow}

\newcommand\ngca{NGC~6334A}
\newcommand\hii{H\,{\small II}}
\newcommand\tcrs{CKR02A} 
\newcommand\supa{$^\mathrm{a}$}
\newcommand{\supb}{$^\mathrm{b}$}
\newcommand{\supc}{$^\mathrm{c}$}
\newcommand{\supd}{$^\mathrm{d}$}

\newcommand{\nodata}{\centering\arraybackslash --}
\newcommand{\kms}   {km~s$^{-1}$}

\newcommand{\jpb}   {$\rm Jy~beam^{-1}$}    

\newcommand{\mo}    {$M_{\odot}$}

\title[The arc-shaped radio source in NGC 6334A]{The arc-shaped radio source at the center of NGC 6334A: Is it a colliding wind region of two young massive stars or the bow shock of a runaway star?}

\author[V. Yanza et al.]{Vanessa Yanza$^{1}$\thanks{E-mail: v.yanza@irya.unam.mx},
Sergio A. Dzib$^{2}$,
Aina Palau$^{1}$,
Luis F. Rodr\'iguez$^{1}$,
Josep M. Masqu\'e$^{3}$,\newauthor
Pedro R. Rivera-Ortiz$^{1,4}$,
and Sac-Nict\'e X. Medina$^{5,2}$
\\
$^{1}$Instituto de Radioastronom\'ia y Astrof\'isica, Universidad Nacional Aut\'onoma de M\'exico, Antigua Carretera a P\'atzcuaro 8701, Ex-Hda. San Jos\'e de \\la Huerta, 58089 Morelia, Michoac\'an, M\'exico\\
$^{2}$Max Planck Institut f\"ur Radioastronomie, Auf dem H\"ugel 69, D-53121 Bonn, Germany\\
$^{3}$Departamento de Astronom\'{\i}a, Universidad de Guanajuato, Apartado Postal 144, 36000 Guanajuato, M\'exico\\
$^{4}$ Escuela Nacional de Estudios Superiores, Universidad Nacional Autonoma de Mexico, Antigua Carretera a P\'atzcuaro 8701, Ex-Hda. San Jos\'e de \\la Huerta, 58089 Morelia, Michoac\'an, M\'exico\\
$^{5}$German Aerospace Center, Scientific Information, D-51147 Cologne, Germany
}

\date{Accepted XXX. Received YYY; in original form ZZZ}

\pubyear{\the\year{}}

\begin{document}
\label{firstpage}
\pagerange{\pageref{firstpage}--\pageref{lastpage}}
\maketitle

\begin{abstract}
New multi-wavelength {\it Karl G. Jansky} VLA observations of \tcrs, the compact radio source in the center of the compact HII region \ngca, are presented. The observations were carried out in five epochs and included the frequency ranges 8.0 - 12.0~GHz (X-band), 18.0 - 26.0 GHz (K-band), and 29.0 - 37.0 GHz (Ka-band). The source is detected and resolved in all the observed epochs and in all bands. The source shows a clear arc-shaped structure consistent with a bow shock. The analysis of the spectral index maps indicates that its spectral index is $\alpha=-0.68\pm0.17$, suggesting that the emission is non-thermal. Two astronomical objects can explain the emission nature and morphology of the source: a colliding wind region of two massive stars or the bow shock of a massive runaway star. However, no massive stars are reported so far in the center of \ngca, though its presence is also suggested by the free-free radio emission of the C-HII region itself. Using ancillary VLA data, we measured a preliminary proper motion of $19\pm6$\,mas\,yr$^{-1}$, equivalent to a velocity of $120\pm40$\,km\,s$^{-1}$.  
A detailed discussion of the implications of both scenarios is provided.
Finally, a list of compact radio sources in the vicinity of \ngca\ is given and briefly discussed.
\end{abstract}




\section{Introduction}

Evidence of compact radio sources\footnote{Radio sources that are unresolved or slightly resolved on sub-arcsecond centimeter observations.} in the very center of compact (C-), and ultra compact (UC-) \hii\ regions have emerged from high angular resolution observations \citep[e.g.,][]{kawamura1998,carral2002,dzib2013w3,masque2017,viveros2023}. Because they are located close to the geometrical center of these \hii\ regions, it was first suggested that these compact radio sources are tracing the ionized winds of the massive star that ionizes the corresponding \hii\ region \citep[][]{carral2002}. However, the predicted flux densities of winds of main-sequence massive stars \citep[see for example Table 3 in][]{dzib2013cyg} are about one order of magnitude lower than the flux densities measured in these compact radio sources \citep[e.g.,][]{yanza2022,viveros2023}, indicating that other emission mechanisms in addition to winds must be at work.

Data emerging from new radio telescopes, such as the {\it Karl G. Jansky} Very Large Array \citep[VLA;][]{perley2011} of NRAO, has started to reveal the nature of these compact radio sources. Instead of a common emission mechanism, the nature of these objects is diverse, such as photo-evaporated disks \citep{dzib2013w3}, magnetically active low-mass stars \citep{dzib2016}, and ionized stellar envelopes \citep{medina2018}. Therefore, it is important to study further C-HII regions to determine if any of these phenomena are predominant during the process of high-mass star formation.

In this work, we present a multi-wavelength study of a particular compact radio source in the center of the NGC~6334A C-HII region.  
This C-\hii\ region was
first reported by \citet{rodriguez1982} and is located at a distance of $1.34\pm0.15$~kpc \citep{reid2014}. It is part of the NGC~6334 complex, a massive star-forming region of 10 pc length composed of several massive clumps \citep[e.g.,][]{cheung2978, gezari1982}.
A shell-like structure was reported in \ngca\ by
\citet{carral2002}, who measured a diameter of $\sim$ 15$''$ and a thickness of $\sim$ 2$''$. 
\citet{carral2002} also found an unresolved radio source near to the geometrical center
of \ngca, hereafter we will refer to this source as \tcrs.
These authors suggested that \tcrs\ probably traces the ionized stellar 
wind of the massive star ionizing the \hii\ region. Ionized stellar winds 
of massive stars are expected to be steady thermal radio emitters \citep{panagia1975}.
However, multi-epoch and multi-wavelength studies by \citet{rodriguez2014}, 
show that the source is variable and has a spectral index ($\alpha$) of $-1.2\pm0.5$
(where $S_\nu\propto\nu^{\alpha}$), properties that point to a radio source with 
non-thermal emission and are consistent with optically-thin synchrotron radiation. 
These authors suggest that the characteristics fit those of known  
strong shocks in the colliding wind regions (CWR) of massive binary stars 
\citep[e.g.,][]{contreras1997b,contreras1999,doug2005,dzib2013cyg}. CWR have an 
arc-shaped morphology as a result of the different wind velocities. Deeper high-angular resolution observations, as those presented in this work, are needed to resolve 
this structure. However, such morphology is also observed in massive runaway 
stars \citep{benaglia2010,rodriguez2020,moutzouri2022}. The velocity of the wind and the motion of the 
massive runaway star generate a speed differential in the direction of its motion, interacting with the surrounding material, and a bow shock is formed.
Because of the strong shock, the radio emission of 
the bow shock is also nonthermal, making the differentiation between the CWR and the runaway star difficult. 
Here, we will explore both scenarios to gain an insight into the nature of \tcrs.

\begin{figure*}
 \centering
    \includegraphics[height=0.5\textheight, trim= 0 10 0 0, clip]{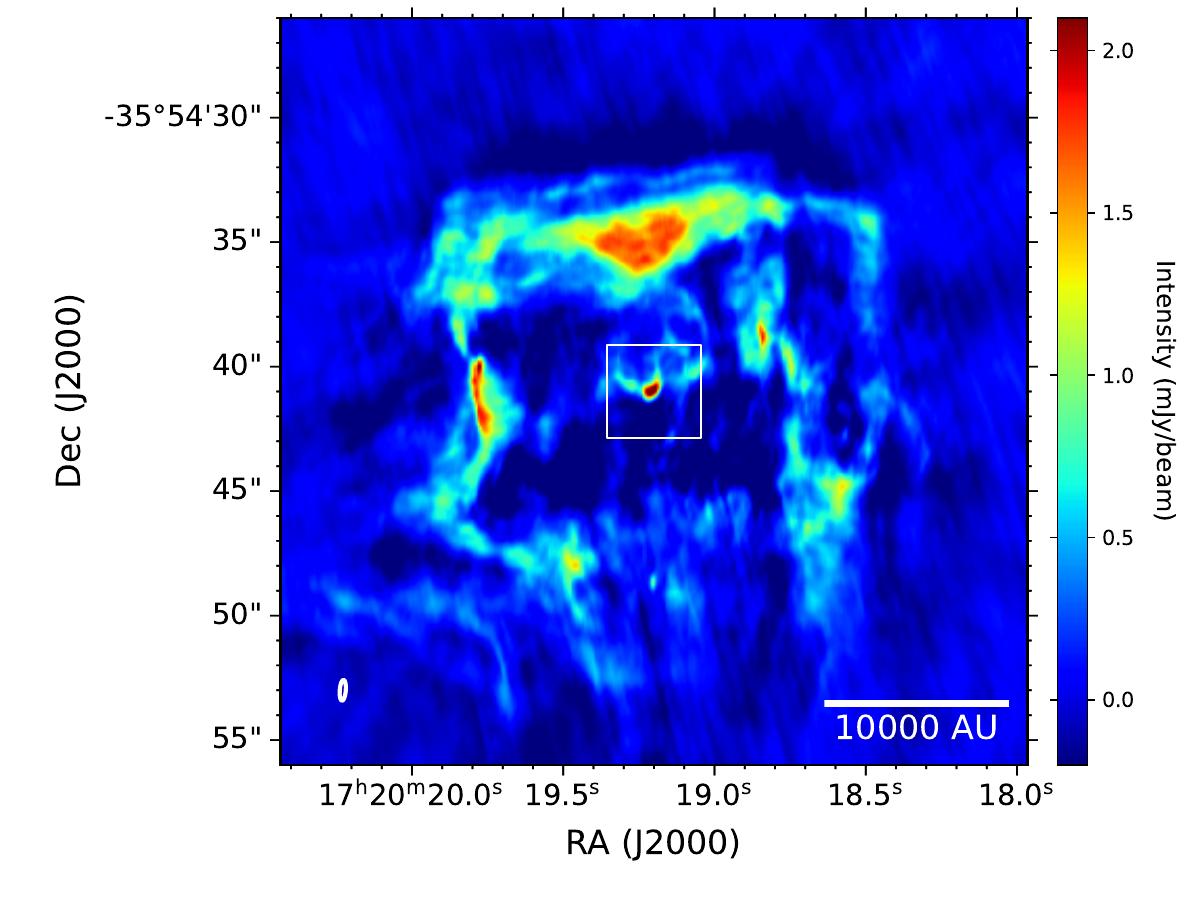}
\caption{Primary beam corrected X band (8-12 GHz) image towards NGC 6334A using all visibilities. The white square indicates the field of view presented in the following figures. The synthesized beam is shown in the bottom-left corner of the panel.} 
\label{fig:square}
\end{figure*}

\section{Observations}\label{sec:obs}

We performed radio observations using the VLA in its A-configuration (project name 14A-014). 
A total of 5 epochs were observed from February 
to April 2014 (see Table~\ref{tab:obs}).  The phase center was $\alpha$(J2000) = 
17$^{\mathrm{h}}$20$^{\mathrm{m}}$19.2$^{\mathrm{s}}$ and $\delta$(J2000)\,
=\,$-35^\circ54'43.0''$. The observations were separated into two groups depending on the spectral setup; 
X-band (8 -- 12 GHz) observations were performed in three epochs, and 
K- (18 -- 26.5 GHz) and Ka-bands (26.5 -- 40 GHz) observations were performed in two epochs.  
The correlator setup on the X-band observations 
covered the full 4.0~GHz bandwidth, while for the K- and Ka-band, 8.0 GHz of bandwidth is covered in each band; 
in these cases, the center frequencies were 22.0 GHz (K-band) and 33.0~GHz (Ka-band). The digital 
correlator of the VLA was configured in 64 spectral windows (SPWs)
for the K- and Ka-bands and 32 SPWs for the X-band. Each SPW bandwidth is 128 MHz divided 
into 64 channels; i.e., a spectral resolution of 2 MHz.

\begin{table}
\small
    \centering
 \caption{VLA radio bservations toward NGC 6334A and image properties.}
    \begin{tabular}{cccccc}
         \hline
         \hline
Epoch & {Date} & {Band} & {Synthesized beam}& {rms noise\supa}  \\
         & (2014) & &  ($\theta_{\rm maj.}\times\theta_{\rm min.}$, P.A.) & ($\mu$Jy\,bm$^{-1}$) \\
\hline
        1\supb& Feb. 24 & K, Ka & ... & ... \\
     2&Feb. 27 & X & $0\rlap{.}''52\times0\rlap{.}''12$, $-07\rlap{.}^{\circ}7$  & 74  \\
      3 &  March 19 & K & $0\rlap{.}''24\times0\rlap{.}''06$,  $-11\rlap{.}^{\circ}9$  & 35   \\
               &       & Ka & $0\rlap{.}''17\times0\rlap{.}''04$,  $-16\rlap{.}^{\circ}3$  & 17   \\
          &       & K+Ka\supc & $0\rlap{.}''18\times0\rlap{.}''05$,  $-15\rlap{.}^{\circ}3$  & 15   \\   
      4 &  April 9 & X & $0\rlap{.}''53\times0\rlap{.}''12$, $+06\rlap{.}^{\circ}5$  & 88 \\
      5 &  April 12 & X & $0\rlap{.}''59\times0\rlap{.}''12$, $-18\rlap{.}^{\circ}9$  & 87 \\
      2+4+5\supd  &     & X & $0\rlap{.}''47\times0\rlap{.}''13$,  $-04\rlap{.}^{\circ}4$ & 170 \\
       & & X  & $0\rlap{.}''40\times0\rlap{.}''12$,  $-06\rlap{.}^{\circ}4$  & 65\\
         \hline
    \end{tabular} \label{tab:obs}
\begin{list}{}{}
    \item[\supa] In the area surrounding the arc-shaped source and avoiding it.
    \item[\supb] Discarded epoch.
    \item[\supc] Image combining the two bands.
    \item[\supd] Image combining the three epochs where the X-band was observed.
\end{list}
\end{table}

In all observations the flux and band-pass calibrator was the quasar J1331+3030 (3C\,286), and the phase calibrator 
was the quasar J1744--3116. The phase calibrator is located at an angular distance of $7\rlap{.}^\circ5$ from the target,
being the nearest acceptable phase calibrator of the VLA calibrator catalog given its flux and structure. 
The first science scan is used to observe the flux calibrator. The following scans observed the target source bracketed with one-minute scans on the phase calibrator. The scan duration on the target source depends on the observed frequency, being 10, 5, and 2 minutes for the X-\,, K-, and Ka-band, respectively.   
For the higher frequencies (K- and Ka-bands), pointing scans at a lower frequency (X-band) are introduced before the first scans on the calibrators, and this is repeated after one hour. The length of the observation is 45 (X-band) and 120 (K- and Ka-bands) minutes. 
A description of all observed epochs is given in Table~\ref{tab:obs}.

The data were calibrated and edited in the standard manner using the Common Astronomy Software Applications \citep[CASA, ][]{mcmullin2007,casa2022}
package of NRAO and the pipeline provided for VLA observations. Imaging was also performed in the CASA software using its {\it tclean} task. However, the process was different for the two spectral setup groups. 

The imaging process for K- and Ka- bands was similar. A Briggs weighting was used with robust = --0.1. Besides, a uv-range $>100\,$k$\lambda$ was applied in these cases. 
Images have sizes of 9000 and 7000 pixels, with square pixel sizes of 0\rlap{.}$''$019 and 0\rlap{.}$''$015 for K- and Ka-bands, respectively, covering square areas with side sizes of 2\rlap{.}$'$85 and 1\rlap{.}$'$75. Finally, a primary beam attenuation correction was applied assuming a Gaussian beam pattern and an attenuation limit of 20\%.  Following the same process, an image was produced by combining the data of these two bands for each epoch. 
After the imaging process, it was found that in one epoch (Feb. 24, K- and Ka-bands, Table~\ref{tab:obs}), the measured fluxes of the sources were systematically 
lower by three orders of magnitude than the other epoch observing the same bands or even than the X band. This indicates an issue in the observations of this epoch, and it was discarded from our analysis.

In the case of the X-band observations, on the other hand, the standard calibration led to maps with lower quality than expected. Because of the presence of the strong extended radio emission from \ngca\ and the nearby bright quasar NGC~6334B \citep{rodriguez1982,moran1990}, the noise levels in the images are higher than the theoretical expected values (for example, rms noise $\sim$ 150 $\mu$Jy beam$^{-1}$ in the image combining the three epochs while the theoretical rms noise is on the order of 10 $\mu$Jy beam$^{-1}$). For these reasons, additional procedures were done. In this band, NGC~6334B has a flux density of $484$ mJy and is only slightly resolved, in principle good for a self-calibration process. However, at an angular distance of $\sim2'$ from the observed phase-center, it falls at the edge of our central primary beam and, then, the primary beam attenuation is different through the observed frequencies of the correlated bandwidth. Thus, to self-calibrate the data, we first divided the data by scans and spectral windows. Then, each data portion was
self-calibrated in phase and amplitude. Additionally, the NGC 6334B cleaned components were subtracted from the visibility set (using the CASA task {\it uvsub}). Finally, the data portions are merged into a single data file again. 

During the imaging process of the X-band, we first made an image combining the three epochs using robust=--0.2 and all the visibilities to recover the shell-like structure of \ngca.
We also made another X-band image (combining the three epochs) using only baselines $>$100 k$\lambda$ and a robust=--0.2, in order to decrease the noise contamination from extended emission and to identify the most compact sources. For this last image, the emission spectrum was modeled as a straight line with a slope in the cleaning process ({\it nterms=2}, in CASA). This spectrum slope is the so-called spectral index ($\alpha$; $S_\nu\propto\nu^\alpha$). 
Table~\ref{tab:obs} includes the synthesized beam and rms noise obtained in the central part of each image.

\section{Results} \label{sec:res}


In Figure \ref{fig:square}, the X-band image of \ngca\ including all visibilities is presented. The image reveals a squared-shell-like structure with \tcrs\ well located in its center, and with the southern side almost absent. This squared structure was already reported by \citet{carral2002}, who observed the region with a synthesized beam three times larger. Thanks to our higher angular resolution, we resolve substructure within each of the main sides of the square reported by \citet{carral2002}. In Sec.~\ref{sec:inter}, the possible interaction of the shell with the surrounding medium is discussed.

In the image generated with $uv$-range $>100\,$k$\lambda$, most of the emission of the square-like structure is filtered out, as expected.  However, some remnants related to the bright edges of the C-HII region are still detected in the final maps, increasing the noise level around \tcrs\ with respect to the noise theoretically expected. 

 We performed a visual inspection combined with an automatic source extraction using the BLOBCAT software \citep{hales2012} to identify radio compact sources in the NGC 6334A region. The visual inspection was done to discard possible artifacts due to residual extended emission, secondary lobes or sources at the limit of the field of view. The radio sources identified with BLOBCAT after the visual inspection are fully described in Appendix \ref{App:CRS}. The Appendix describes the details of the compact radio source population detected in the field of view of our observations at the different frequencies and their properties. In what follows, this work will focus on understanding the nature of the CKR02A central radio source.

\subsection{Structure and emission properties of \tcrs}

\begin{figure*}
 \centering
    \includegraphics[height=0.5\textheight, trim= 0 0 0 0, clip]{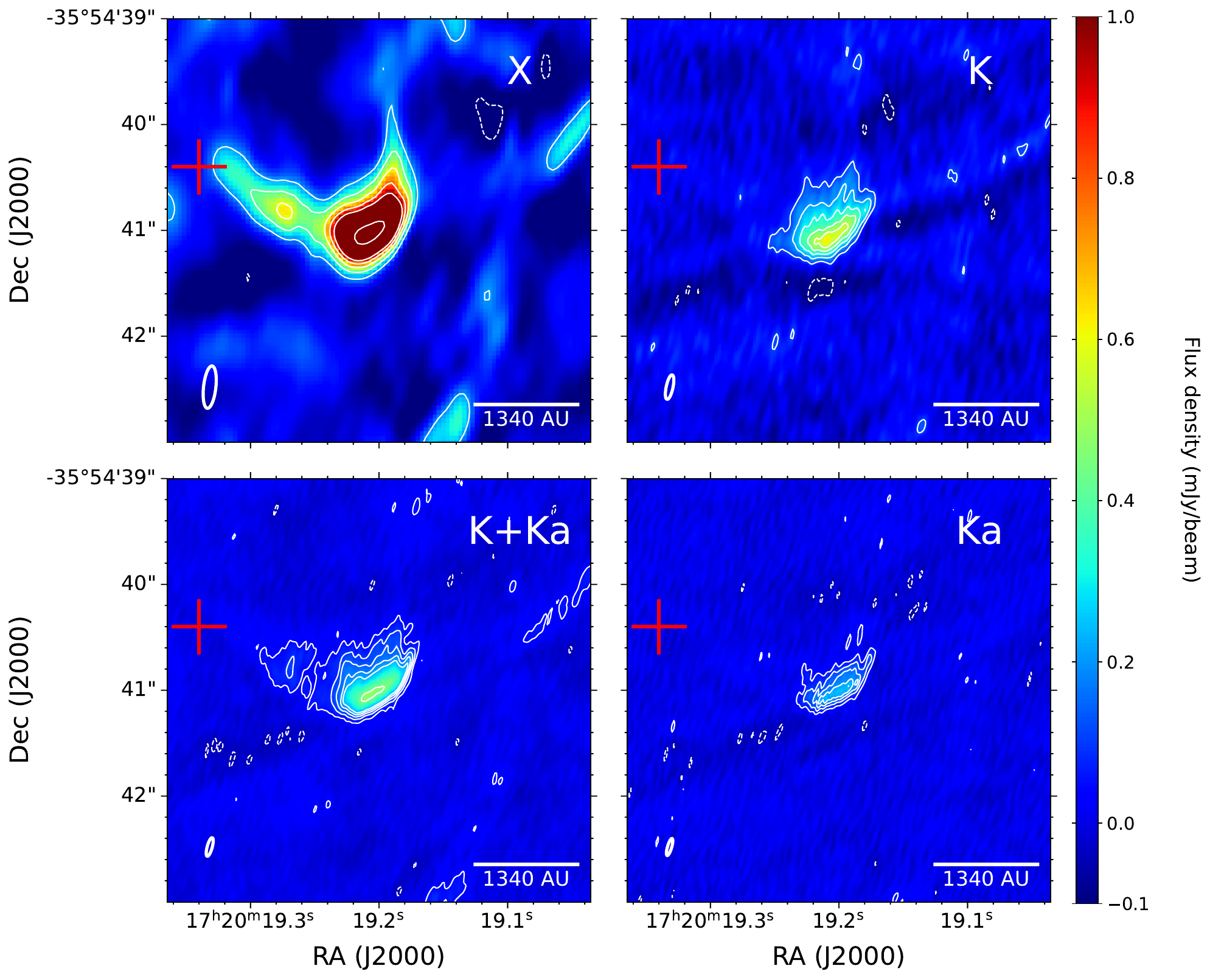}
\caption{Radio map at different bands of the CKR02A at the center of the NGC 6334A region. All images were made with uv-range$>$100k$\lambda$. Panels are: (a) X-, (b) K-, (c) K+Ka-, and (d) Ka-bands.
Contour levels are -3, 3, 6, 9, 12, 15, and 30 times the noise level in the map, as listed in Table~\ref{tab:obs}.
The synthesized beam size of each map is shown as a white ellipse in the bottom-left corner of the corresponding panel. The red cross indicates the position of the Class I young stellar object identified by \citet{willis2013}, with the size corresponding to half of the beam of the near-infrared observations of this work (see main text in Sec.~4).}
\label{fig:CWR}
\end{figure*}

\tcrs\ is the third brightest source detected in the region. It is detected in all observed bands, see Fig.~\ref{fig:CWR}. On inspection of Fig.~\ref{fig:CWR}, it
is noted that \tcrs\ has an arc-shaped structure at all observed bands.  The main position of the peak pixel in \tcrs\ is $\alpha$(J2000) = 
17$^{\mathrm{h}}$20$^{\mathrm{m}}$19.205$^{\mathrm{s}}$ $\pm$ 0.001$^{\mathrm{s}}$ and $\delta$(J2000)
= $-35^\circ54'41.04''$ $\pm$ 0.01$''$. It is also noted that its
brightness is higher at lower frequencies. 
To obtain the source flux density and peak intensity, we measured them in the image with $uv$-range$\,>$100k$\lambda$ within the 3$\sigma$ contour level, where $\sigma = 56\,\mu$\jpb. We fitted a 2D Gaussian to measure the size. These results are listed in 
Table~\ref{tab:2properties}.

\begin{figure*}
 \centering
    \includegraphics[height=0.5\textheight, trim= 0 10 0 0, clip]{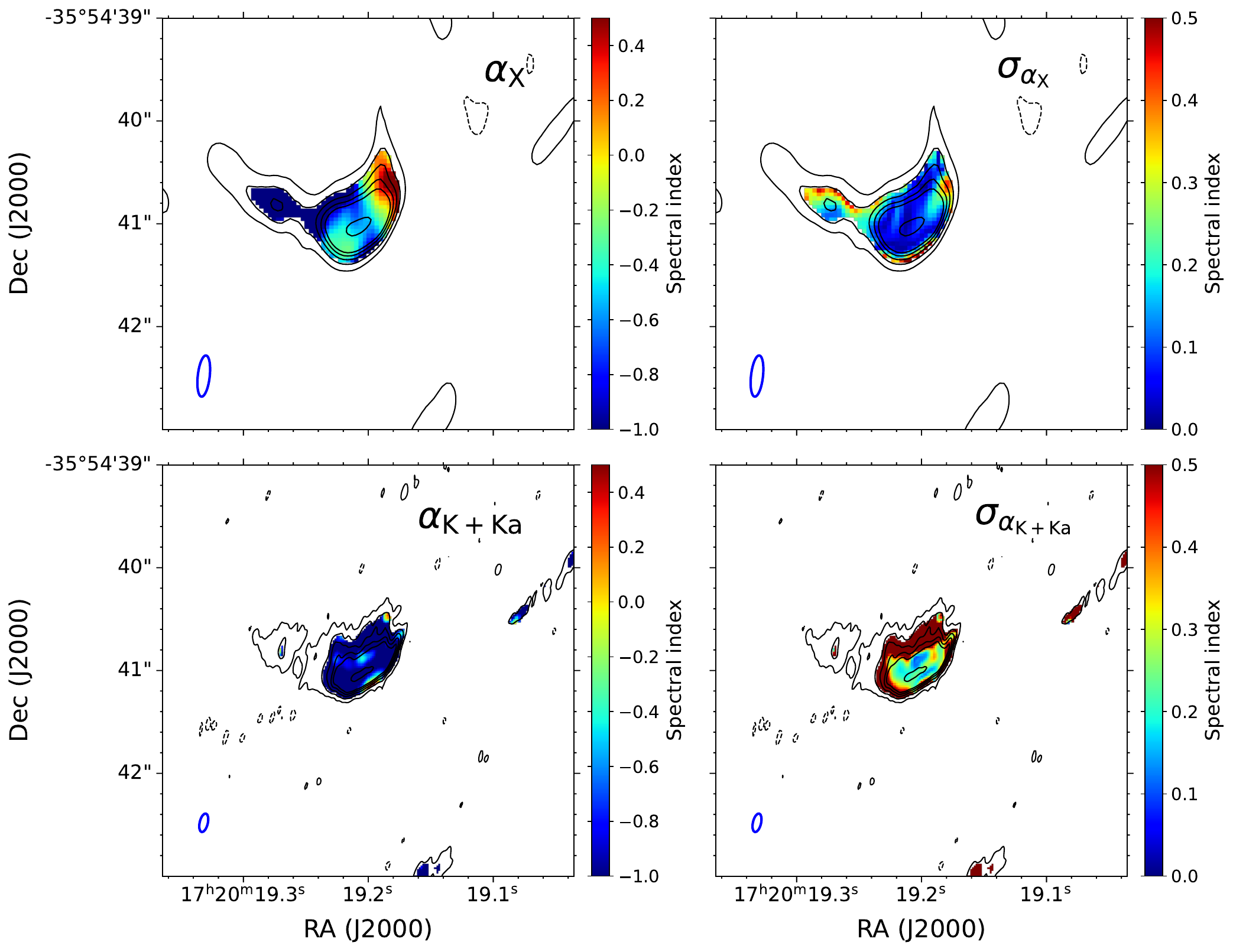}
\caption{Spectral index maps of \tcrs\ obtained using nterms=2 in the clean task of CASA with uv-range$>100~k\lambda$ and $100~k\lambda<$ uv-range $<1630~k\lambda$ for X and K+Ka bands respectively. Panels are: (a) Spectral index map obtained from all observations in full X-band, (b) spectral index error in the X band, (c) spectral index from the combination of the K- and Ka-bands, and (d) spectral index error in the K+Ka band. Spectral index results are given in the regions where the radio continuum is  $>6\,\sigma_{\rm noise}$. Regions below this value are masked. Contour levels and synthesized beam sizes are the same as in Fig.~\ref{fig:CWR} for their corresponding band. 
} 
\label{fig:alpha}
\end{figure*}

\subsection{Spectral index}

On a logarithmic scale, the flux density at radio frequencies has a linear shape with the form: 

\begin{equation}
\centering
    \log{S_\nu}=\alpha\cdot\log{(\nu)}+C,
\label{eq:alpha}
\end{equation}

\noindent where $\alpha$ and C are the slope (corresponding to the spectral index) and the intercept, respectively. 
As a first estimation of the spectral index value, we used the measured flux 
densities in the three observed bands (X, K, and Ka) reported in Table \ref{tab:2properties}, and performed a 
least squares fit to these values, resulting in $\alpha=-0.52\pm0.03$.


This result can be compared with a spectral index map obtained using nterms=2 in the clean task of CASA (Section \ref{sec:obs}).
Two cases were considered for the analysis. First, the combination of the three
epochs where the X-band was observed as representative of the low-frequency 
case and, second, the combination of the K and Ka band data as representative
for the high-frequency case. Since the variability reported in the source is of the order of years \citep{rodriguez2014}, we expect that variability does not affect the X-band flux in our different epochs that are separated by only one month and a half. We verified that variability is not present in our data, by measuring the X-band flux densities for each epoch, finding changes within 1$\%$. Therefore, we consider that the combination of the three epochs provides a good estimation of the spectral index and used these epochs to reach a higher sensitivity. For the K and Ka bands, we generated an image using only the common uv-coverage of the two bands. As we have already limited the uv-range to $>$100~k$\lambda$ to filter out the extended emission and make the cleaning process easier (Section \ref{sec:obs}), we used the same lower limit to obtain the spectral index map, i.e. we used a uv-range between of 100--1630~k$\lambda$.

\begin{figure}
 \centering
    \includegraphics[width=0.5\textwidth, trim= 0 65 0 100, clip]{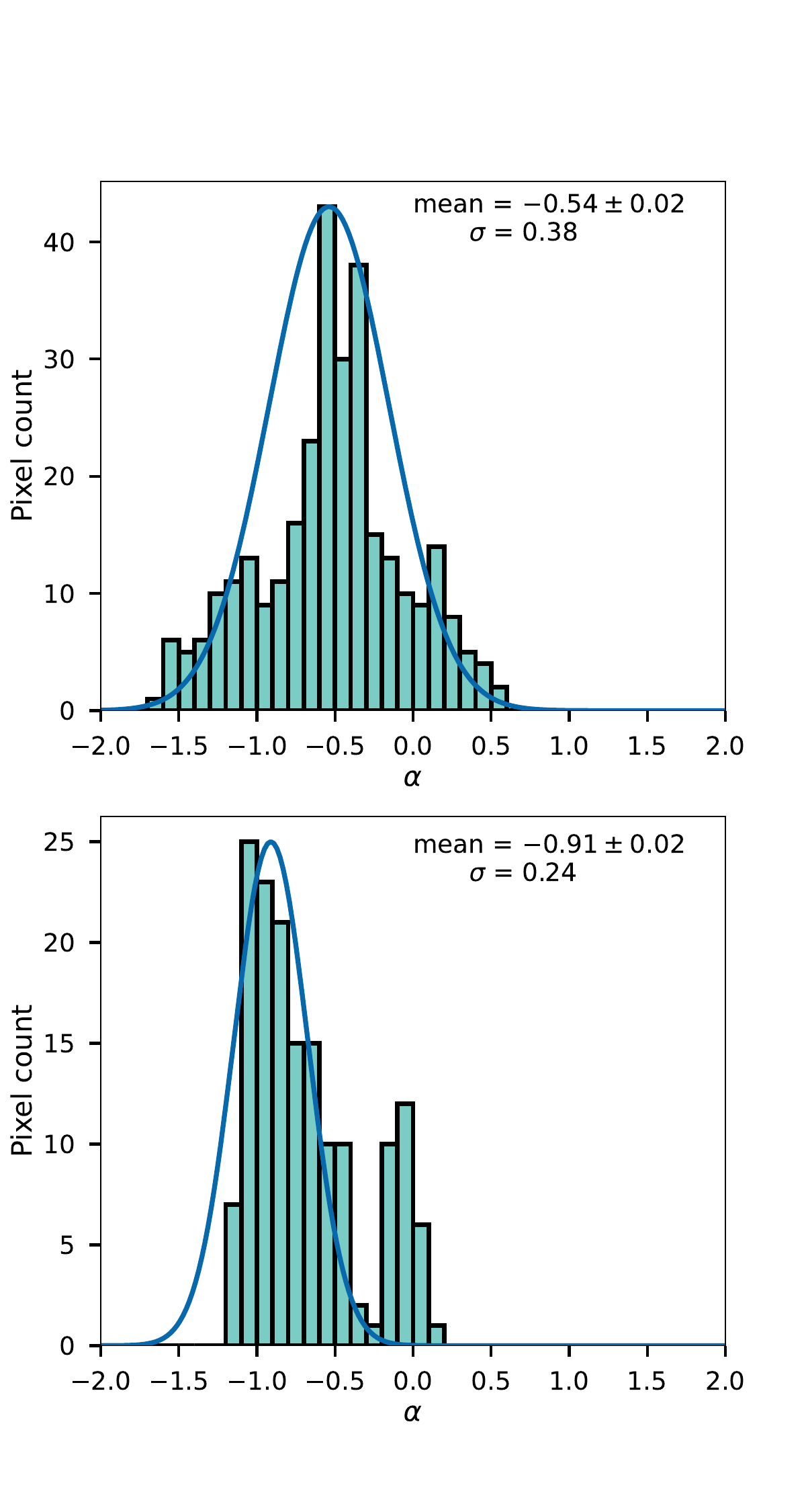}
\caption{Distribution of spectral index pixels. {\it Top:} Pixels from the spectral index map of the X-band image. {\it Bottom:} Pixels from the spectral index map of the K+Ka band image. }
\label{fig:aHist}
\end{figure}

The two maps and their corresponding errors are shown in Figure~\ref{fig:alpha}.
In these maps, the spectral index values range from $-1.0$ to 0.5, and the errors from 0.0 to 0.6. To estimate
the mean value of these pixels, first we only kept those with good determination
and removed all pixels whose error is $>0.2$. Second, in Fig.~\ref{fig:aHist}, histograms of pixel value distributions are shown. Finally, we fit these
histograms by using a Gaussian function. The mean spectral indices are 
$\alpha_{\rm X}=-0.54\pm0.02$ and $\alpha_{\rm K+Ka}=-0.91\pm0.02$, with standard deviations of $\sigma_{\rm X}=0.38$ and $\sigma_{\rm K+Ka}=0.24$. We performed a weighted average of
these two results, assuming the standard deviations as the final errors, and found $\alpha=-0.68\pm0.17$ for \tcrs.

\begin{table}
\footnotesize
\renewcommand{\arraystretch}{1.0}
\setlength{\tabcolsep}{3pt}
    \centering
 \caption{ \tcrs\ properties obtained from the images with the same $uv$-range shown in Fig. \ref{fig:alpha}.}
    \begin{tabular}{c c c c c c}
         \hline
         \hline
        Band & Size\supa &  $S_{\nu, {\rm Int.},3\sigma}$\supb &  $I_{\nu, {\rm Peak}}$ \\
         &   ($\theta_{\rm maj.}['']\times\theta_{\rm min.}['']$; P.A.$[^{\circ}]$) & (mJy)   & (mJy\,bm$^{-1}$) \\
         \hline
         
X\supc   & $0.60\pm0.07\times0.29\pm0.09$; $119\pm10$& \, $15.8\pm0.4$ & $2.16\pm0.07$\\
K   &$0.60\pm0.05\times0.37\pm0.04$; $120\pm9$& \, $8.0\pm0.2$ & $0.62\pm0.04$\\
Ka  &$0.50\pm0.03\times0.23\pm0.02$; $124\pm4$& \, $3.8\pm0.1$  & $0.25\pm0.02$\\
K+Ka\supd&$0.57\pm0.03\times0.30\pm0.02$; $121\pm4$& \, $10.21\pm0.1$ & $0.47\pm0.02$\\
         \hline
   \end{tabular}
\begin{list}{}{}
    \item[\supa] The size is measured from a Gaussian fit, taking the deconvolved value.
    \item[\supb] The flux density is measured within the 3$\sigma$ contour. The error is calculated following Appendix A2 of \citet{Beltran2001}.
    \item[\supc] From the image combining the three epochs where the X-band was observed.
    \item [\supd] From the image combining the two bands.
\end{list}
 \label{tab:2properties}
\end{table}

\section{Discussion}

\tcrs\ was detected by \citet{rodriguez2014}, who first suggested that this source 
is a non-thermal emitter. They also suggested that its emission properties better 
fit those of wind-wind collision regions of massive binary stars. However, because 
of the limited angular resolution of their observations the source was only slightly 
resolved, and no well-defined structure was identified. 

In the wind-wind collision process, the winds of the two massive stars strongly interact producing a front shock with an arc-shaped bow structure \citep[e.g., ][]{canto1996,doug2005,dzib2013cyg}. In the shocked 
region electrons are accelerated to relativistic velocities producing optically-thin synchrotron 
radio emission with a typical spectral index of --0.5 \citep[e.g.,][]{doug2003}.
With the new images presented in this work, we have, for the first time, revealed that 
the morphology of \tcrs\ is that of a bow shock. Our spectral index analysis also confirms that it is a non-thermal radio source, as its spectral index is negative 
($\alpha=-0.68\pm0.17$). 

In star-forming regions, four main types of compact 
radio sources could have an arc-shaped morphology: i) UC-HII regions \citep[e.g., G34.26+0.15,][]{churchwell2002}; ii) proplyds \citep[e.g.,][]{zapata2004,forbrich2016}; iii) CWRs \citep[e.g.,][]{dzib2013cyg} and 
iv) massive young runaway stars interacting with the surrounding material \citep[e.g.,][]{rodriguez2020}. \tcrs\ is discarded to be an UC-HII region or a 
proplyd, as the radio emission from these objects is thermal. The same argument applies to discard a photoevaporating disk \citep[e.g.,][]{Hollenbach1994}.
Bow shocks of CWRs or runaway stars, on the other hand, can be
non-thermal radio emitters \citep[e.g.,][]{doug2003,dzib2013cyg,benaglia2010,moutzouri2022}. 
In the following subsections, we will describe these two possibilities.

It is important to note that no evidence of the putative massive star(s),
which are required to explain the bow shock in the two possible scenarios (CWR or runaway star), has
been found so far. The nearest Spitzer (IRAC) sources are located $\sim5''$ to
the southeast or $\sim10''$ to the northwest \citep[e.g.][]{Tige2017} of CKR02A. The
infrared source closest to CKR02A is reported by \citet{willis2013} and was
detected with the National Optical Astronomy Observatory (NOAO) Extremely Wide-Field Infrared Imager (NEWFIRM) camera on the Blanco 4\,m telescope at Cerro Tololo Inter-American Observatory. The position of this source is indicated in Fig.~\ref{fig:CWR}. This source, SSTU J172019.34$-$355440.4, is classified as a Class I young stellar object and lies about $2''$ to the northeast of CKR02A. Since this infrared source is fainter than the completeness limit estimated for IRAC1 for Class I
sources, which corresponds to objects of about 1~\mo\ \citep{willis2013}, it must be of low-mass. Therefore, no evidence of high-mass stars are found so far associated with CKR02A at optical or infrared wavelengths.

\subsection{CWR scenario}

Because fast and very strong winds are needed to produce CWRs with non-thermal 
radio emission \citep[e.g.,][]{doug2003}, these are usually mainly associated with
evolved massive stars losing mass through their strong winds, such as Wolf-Rayet (WR) and Luminous Blue Variable stars \citep[e.g., ][]{doug2003,doug2005,parkin2011}. 
They have also been detected in main sequence massive binary systems that have cleared most 
of the HII region where they were born \citep[e.g.,][]{dzib2013cyg,sb2019}. To our knowledge,
CWRs have not been detected in embedded massive binary systems. Opacity from 
the thermal free-free emission of young HII regions may play an important 
role in preventing the detection of CWRs in young binary systems as the non-thermal emission could be absorbed, and also the larger flux density of the HII region will make the detection of the relatively weaker CWR difficult.

\subsubsection{Stellar components} \label{sec:stellarcompo}

No massive stellar sources have been reported, at any wavelength, to be 
located in the center of \ngca. However, to ionize the C-HII region,
it is inferred that at least one massive star must reside in its interior. 
Initially, \citet{rodriguez1982} suggested that an O7.5 star is ionizing \ngca.

Considering a better distance estimation and flux density from more recent observations, 
we estimated the spectral type of the star by measuring the rate of ionizing photons, following \cite{sanchez-monge2013}:

\begin{equation}
    \left[ \frac{\dot{N}_{i}}{{\rm s}^{-1}} \right] = 8.852 \times 10^{40}\left[ \frac{S_{\nu}}{\rm Jy} \right] \left[ \frac{\nu}{\rm GHz} \right]^{0.1}  \left[ \frac{T_{e}}{10^{4} {\rm K}} \right]^{0.35} \left[ \frac{D}{\rm pc} \right]^{2},
\label{eq:ni_photons}    
\end{equation}

\noindent where $S_{\nu}$ is the flux density, $T_{e}$ is the electron temperature, $\nu$ is the frequency and $D$ is the distance to the region. 
The flux density of the square shell was measured from the VLA project 14A-241, at 23.2 GHz and using the D configuration. These data are sensitive to more extended emission associated with the ionized shell. For this case, $S_{\nu} = 9$~Jy, and we assumed $T_e =$ 5000--10000~K and $D = 1340$~pc. The ionizing photon rate is $\dot{N_{i}} =$ (1.5--2.0)$\times 10^{48}$~s$^{-1}$, corresponding to a B0--O8 star, following \citet{panagia1973} and \citet{vacca1996}.  

To produce a CWR with non-thermal radio emission, a second massive star 
with strong winds must exist in the interior of \ngca. This second star should be of spectral type 
B0 or earlier, to produce strong winds that interact with those of the B0--O8 star. 

Assuming that both stars are in the main sequence, the strength of their winds correlates with their masses. The orientation of the bow shock structure with the tip pointing southward suggests that the most massive star is located to the southwest of \tcrs\ and the less massive star to the northeast.    

Stellar winds of massive stars are made of ionized gas, and thus  they are a source of thermal
emission. Besides \tcrs\ there are no other compact radio sources in the 
inner parts of \ngca\ which could point to the position of these massive 
stars. This, however, is not surprising. \citet{dzib2013cyg} estimated 
the expected fluxes from ionized winds of massive stars located at 1 kpc,
at an observed frequency of 8.46 GHz (values listed in their Table~3) and find that for typical massive 
stars (B0 to O7) they are rather weak ($<0.1$~mJy at 1.0~kpc). At a distance 
of 1.34~kpc, these should be weaker and at levels comparable to the image 
noise or below. Next-generation interferometers such as the ngVLA that will provide both
high angular resolution and higher sensitivity will be capable of detecting the radio emission of the wind of these stars, if present.

\subsubsection{Separation between the two stellar components} \label{sec:separation2binary}

In the CWR scenario and assuming that the involved stars are main sequence O8 and B0  stars (Sec.~\ref{sec:stellarcompo}),  approximated positions of the stellar components can be
estimated. We will use the formulation derived by \citet{eichler1993} for CWRs of massive stars.
First, the angular separation of the massive star with the less powerful wind (the B0 star) to the CWR, $r_{B}$, was calculated using $r_{B} \approx l^{rad}/\pi$
\citep[equation 19 in ][]{eichler1993}. In this equation, $l^{rad}$ is the characteristic size of the non-thermal radio emission in the CWR, i.e., the length of the bow shock. Given the morphology of \tcrs, a Gaussian fit is not appropriate. Then, we took two different values as a maximum and a minimum limit for the length, $l^{rad} = 0\rlap{.}''6$ and $l^{rad} = 1\rlap{.}''8$ (measured from the image). Thus, we obtained a range of possible angular separations for the less-massive stellar component of $r_{B-1}= 0\rlap{.}''2$ and $r_{B-2}= 0\rlap{.}''6$, respectively, to the north-east from the bow shock. We note that SSTU J172019.34$-$355440.4 is located $\sim2''$ to the north-east of the bow shock (see Fig.~\ref{fig:CWR}), but it is a low-mass young stellar object and it is difficult to reconcile with the required massive nature of the stars producing a CWR.

\citet{eichler1993} showed that in the case of two spherical winds, the distance of the star with the most powerful wind (the O8 star) to the bow shock, $r_O$, is related to the distance to the less powerful wind star (their equation 1) $r_B$ as:

\begin{equation}
    r_O=\frac{1}{\eta ^{1/2}} r_B, 
    \label{eq:ro}
\end{equation}

\noindent where $\eta$ is the ratio of the momentum of the stellar winds. Then, $\eta$  is defined as:
\begin{equation}
\centering
  \eta= \frac{ (\dot{M}_{B} \times v_{\infty B})}{(\dot{M}_{O} \times v_{\infty O})}.
  \label{eq:eta}
\end{equation}

We used the mass loss rate and wind velocity reported by \citet{vacca1996} and summarized by \citet{dzib2013cyg}. 
For the O8 star, $\dot{M}_O=10^{-6.24}$ $M_\odot$~yr$^{-1}$ and $v_{\infty O} = 2600$~km~s$^{-1}$.
In the case of the B0 star, $\dot{M}_B = 10^{-6.99}$ $M_\odot$~yr$^{-1}$ and $v_{\infty B}= 2500$~km~s$^{-1}$. Using these values, we find that $\eta=0.17$. 
Finally, using equation \ref{eq:ro}, the angular separation of the most massive stellar component ranges from $r_{O-1} = 0\rlap{.}''48$ to $r_{O-2} = 1\rlap{.}''45$ to the southwest from the bow shock.

\subsection{The runaway star scenario}\label{sec:runaway}

Alternatively, the bow shock could be interpreted as produced by a runaway star. In this case, the bow-shock morphology indicates the trajectory of the star. For the case of \tcrs, the bow shock shape suggests that the star is moving from the north(east) to the south(west). In this scenario, the non-thermal radio emission arises from the interaction of the stellar wind of the star with the interstellar medium. \citet{benaglia2010} and \citet{moutzouri2022} report bow-shocks from massive runaway stars with non-thermal spectral indices. This scenario will be explored below.

\subsubsection{Velocity estimations} \label{sec:vels}

VLA observations are known to be a good tool for astrometry. This is because the high angular resolution achievable in the extended configurations and because the observations are phase referenced with quasars, which can be assumed as non-moving sources providing good rest reference points. However, the observations to \ngca\ are far from ideal astrometric experiments. First, the source is at a declination of $\sim-36^\circ$, implying very low elevation observations with the VLA that introduce significant systematic position errors \citep[e.g.,][]{rodriguez2014}. Second, existing archived VLA observations do not use the same phase calibrator, making difficult to compare them. Fortunately, in the field of view of low-frequency observations ($\leq12$\,GHz) the quasar NGC\,63334B is also detected, and we can anchor the astrometry of \tcrs\ to it.  

The velocity of the possible runaway star was calculated by comparing the relative angular separation between the quasar NGC\,6334B, to the north, and the compact source \tcrs, at different epochs; this allows to measure the position of \tcrs\ with respect to a rest reference source. The analysis was made using VLA archive data from several projects at similar configurations (listed in Table~\ref{tab:data}). 
These ancillary data were calibrated, edited, and imaged using standard procedures with the software
Astronomical Image Processing System \citep[AIPS,][]{greisen2003}.
For each epoch, a 2D-Gaussian fit was applied for both the quasar and \tcrs\ to obtain the position of each source and, thus, the angular separation. The error for the angular separation was estimated from the propagation of errors of the uncertainty in the measurement of the positions, and the systematic error estimated from the 1997 epoch in which two different bands were observed.

\begin{figure*}
 \centering
    \includegraphics[height=0.4\textheight, trim= 0 10 0 0, clip]{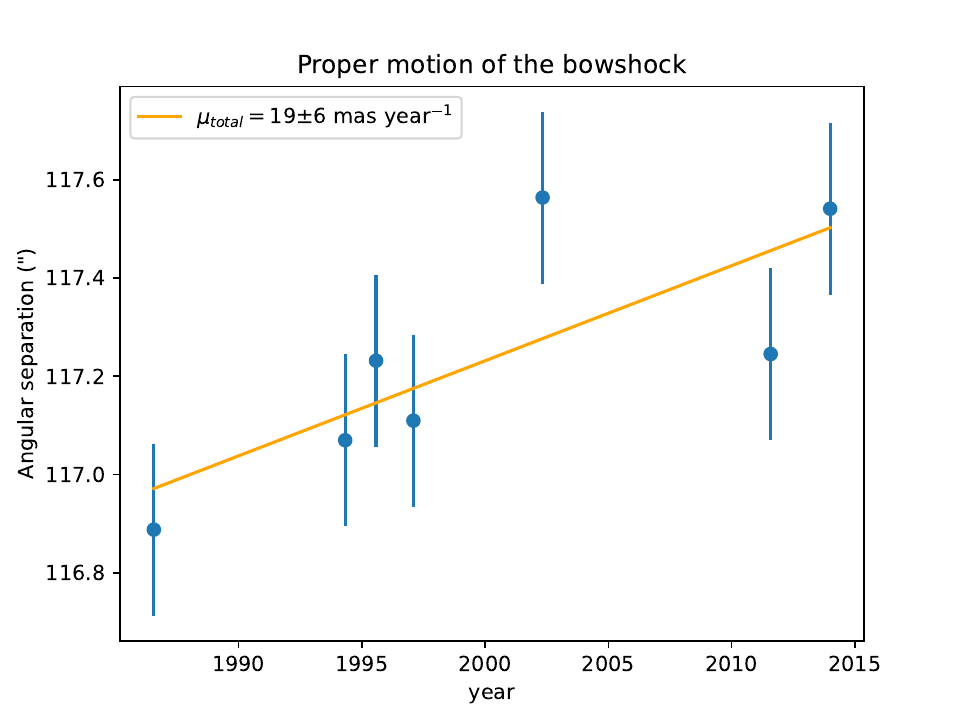}
\caption{Relative angular separation between NGC 6334B and \tcrs\ for different epochs. The yellow line corresponds to the least square fit.} 
\label{fig:relative_sep}
\end{figure*}

The proper motion of the source is presented in Figure~\ref{fig:relative_sep}. This figure shows the relative angular separation for each epoch and their least squares fit.
A positive correlation is apparent in Fig.~\ref{fig:relative_sep}, 
revealing a motion of \tcrs\ with respect to the quasar.
Using the slope obtained in the least squares fit, the proper motion was calculated $\mu_{tot} = 19 \pm 6$~mas\,yr$^{-1}$, which corresponds to a velocity of $v_{tot} = 120 \pm 40$~km\,s$^{-1}$ assuming the distance of 1.34 kpc. Given the significance (3$\sigma$) of our velocity determination, we cannot conclusively discard or confirm the runaway scenario. Further observations providing a broader time baseline would be very helpful to confirm these results.

\begin{table*}
    \centering
    \caption{Data used to explore the proper motion of \tcrs}
    \begin{tabular}{c c c c c c c}
    \hline \hline
         Epoch & Project & Configuration & Central frequency (GHz) & $uv$ cut\supa~(k$\lambda$) & Synthesized beam & Angular separation\supb~($''$) \\
         \hline
         1986/06/23 & AW147 & BnA & 4.83 & 30 & $1\rlap{.}''29\times0\rlap{.}''94$,  $40\rlap{.}^{\circ}5$ & 116.89 $\pm$ 0.09 \\
         1994/04/28 & AM447 & A & 4.86 & 20  & $1\rlap{.}''74\times0\rlap{.}''33$,  $-02\rlap{.}^{\circ}7$ & 117.07 $\pm$ 0.09 \\
         1995/07/29 & AM496 & A & 4.82 & 20 & $1\rlap{.}''85\times0\rlap{.}''36$,  $08\rlap{.}^{\circ}6$ &  117.23 $\pm$ 0.02\\
         1997/02/02 & AT202 & BnA & 8.46 & 20 & $1\rlap{.}''78\times0\rlap{.}''79$,  $40\rlap{.}^{\circ}6$ & 117.11 $\pm$ 0.09 \\
         2002/05/01 & AR474 & A & 8.46 & 40  & $0\rlap{.}''57\times0\rlap{.}''20$,  $-11\rlap{.}^{\circ}4$ & 117.56 $\pm$  0.03\\
         2011/08/02 & SC0113 & A & 8.46 & 20 & $0\rlap{.}''77\times0\rlap{.}''16$,  $-19\rlap{.}^{\circ}1$ & 117.24 $\pm$  0.09 \\
         2014/03/21\supc & 14A-014 & A & 10 & 100 & $0\rlap{.}''42\times0\rlap{.}''13$,  $-05\rlap{.}^{\circ}8$ & 117.54 $\pm$ 0.09\\
         \hline
    \end{tabular}
\begin{list}{}{}
\centering
    \item[\supa]  The images are obtained using visibilities greater than this value.
    \item [\supb] The errors are obtained from the propagation of the errors of the gaussian-position fit.
    \item [\supc] For this epoch a new X-band image was generated combining the three epochs where the quasar visibilities were not subtracted.
\end{list}

    \label{tab:data}
\end{table*}

\subsubsection{Approximated position of the runaway star}

Similar to the analysis developed in section \ref{sec:separation2binary} for the separation between the two stellar components in the CWR scenario, we estimated the distance where the runaway star should be located relative to the bow shock. To do this, equation \ref{eq:runwypos} from \citet{wilkin1996} was used:

\begin{equation}
    R_0 = \sqrt{ \frac{\dot{M}_O v_{\infty}}{4 \pi \rho_a V_{*}^{2}}}\, ,
    \label{eq:runwypos}
\end{equation}

\noindent where $\dot{M}_\mathrm{O}$ is the mass-loss rate of the star, $V_{\infty}$ is the velocity of the wind, $\rho_a$ is the density of an uniform medium and $V_{*}$ is the constant velocity of the star. Assuming an O8 star, $\dot{M}_\mathrm{O}$ and $v_{\infty}$ have the same value taken in section 
\ref{sec:separation2binary}, $\dot{M}_ \mathrm{O}~= 10^{-6.24}$ $M_\odot$~yr$^{-1}$ and $v_{\infty O} = 2600$~km~s$^{-1}$. For the density of the medium, the value of  $10^{5}$~cm$^{-3}$ was adopted (see Appendix~\ref{App:coolingshell}). 
Assuming a mean molecular weight of 2.3, the aforementioned density corresponds to $\rho_a$ = 3.8 
$10^{-19}$\,g\,cm$^{-3}$. Using the velocity estimated in Sec.~\ref{sec:vels}, of 120 km~s$^{-1}$, and assuming that the motion of the star is in the plane of the sky,
we obtained $R_0 = 0\rlap{.}''45$ or $\sim600$~au. 

\subsection{Interaction of \tcrs\ with the surrounding molecular gas}\label{sec:inter}

\begin{figure*}
 \centering
    \includegraphics[height=0.28\textheight, trim= 0 10 0 0, clip]{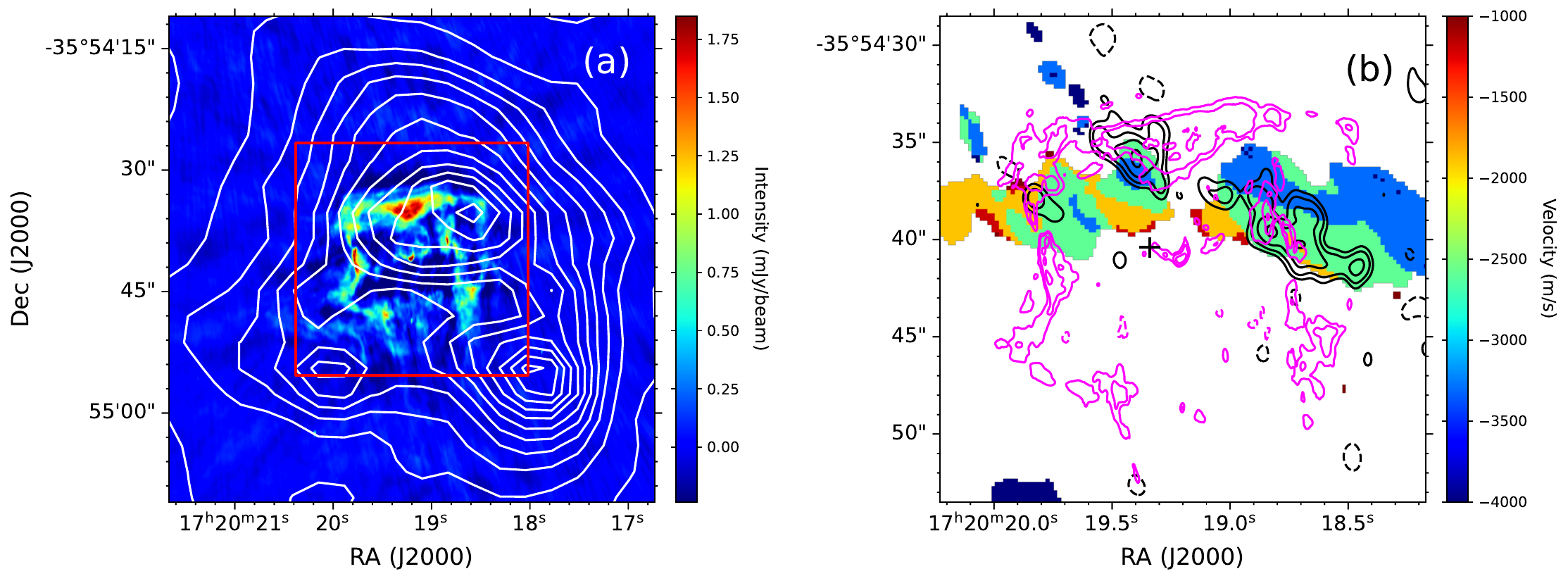}
\caption{ Comparison of our data with the surrounding medium using data previously published. (a) Background image is our data shown in Fig. \ref{fig:square}. The contour image is data from {\it Herschel}, PACS at 70 $\mu m$ used by \citet{Tige2017}. The contour levels are $-3$, 3, 6, 9, 11, 13, 15, 17, 19, 21, 23, 25 and 27 times the rms noise that is 3~Jy~beam$^{-1}$. The red square shows the field of view of the b panel. (b) Background image is the first-order moment of H$^{13}$CO$^{+}$ (4--3) (346.998344~GHz) taken by \citet{palau2021}. The magenta contours show the VLA image at X band (8-12~GHz) using all visibilities presented in Fig. \ref{fig:square}. The contour levels are -3, 3, 5, 10, 20 and 40 $\times$ 0.17~mJy~beam$^{-1}$. Black contours show the 0.87~mm map taken with the SMA by \citet{palau2021}. The black-contour levels are $-4$, 4, 8, 16, 32 times 6.6~mJy~beam$^{-1}$. The black cross indicates the position of the Class I young stellar object identified by \citet{willis2013}, with the size corresponding to the beam of the near-infrared observations of this work (see main text in Sec.~4).}
\label{fig:2contour}
\end{figure*}

Since the only counterpart of the central radio source is a low-mass Class I young stellar object and no massive stars have been found associated with CKR02A, we consider here the possibility that the stars are obscured by the surrounding dust. To explore this, in the left panel of Figure \ref{fig:2contour}, we present the centimeter emission of this work (Fig.~\ref{fig:square}) along with the 70~$\mu$m PACS/{\it Herschel} emission superimposed. The {\it Herschel} emission consists of three main cores that lie at the outer border of the squared shell, with the strongest one located about $10''$ to the northwest of \tcrs, and the other two cores located to the south. The overall {\it Herschel} emission is reminiscent of a shell and \tcrs\ lies in the northern part of the {\it Herschel} cavity.

In the right panel of Fig.~\ref{fig:2contour}, we present the centimeter emission of this work (Fig.~\ref{fig:square}), with the H$^{13}$CO$^{+}$ (4--3) first order moment in color scale and the dust continuum emission at 0.87 mm in black contours, obtained with the SMA by \cite{palau2021}. The figure shows that the dust continuum emission has a filamentary morphology with a northeast-southwest orientation, and that several condensations of the filament coincide in projection with the centimeter shell. The western side of the millimeter filament lies about $5''$ to the south of the {\it Herschel} peak. This millimeter filament matches the large-scale filament orientation of the entire NGC\,6334 cloud \citep[e.g.,][]{arzo2021}. Furthermore, the molecular gas also presents a filamentary morphology with an east-west orientation. While the western side of the molecular filament matches the dust emission, the eastern side lies to the south of the centimeter and millimeter continuum emission.
As we can see from the figure, the molecular gas is located in between \tcrs\ and the northern side of the squared shell. The velocity field of the molecular gas is particularly redshifted near \tcrs, and the velocity progressively becomes more blueshifted to the north. These velocities of the molecular gas, ranging from $-3$ to $-1$~\kms, are very close to the velocities measured for the
ionized gas by \citet{depree1995} using radio recombination lines, which range from $-4$ to 0~\kms. Overall, this suggests a possible interaction between the central source of the squared shell and the surrounding gas. 

From these figures, it seems that \tcrs\ lies near the center of a cavity whose edges are emitting from centimeter to submillimeter wavelengths. Therefore, the center of the cavity where \tcrs\ is located cannot be severely obscured. 

\subsection{\tcrs: Is it a CWR or the bow shock of a runaway star?}

In previous sections, two different possibilities for the nature of \tcrs\ were explored: a CWR or a bow shock of a runaway star. Both scenarios can explain the spectral index and morphology of \tcrs. Given the large-scale cm emission, the first suggestion is that a massive star is ionizing the C-HII region. In our case, as it is explained in section \ref{sec:stellarcompo}, an O8 star is required to achieve the flux density of the extended emission of the square shell. However, there are no massive counterparts detected towards the interior of \ngca. Now, we will discuss the advantages and disadvantages of the two scenarios. 

The first scenario, where \tcrs\ is a CWR, explains the ionization of the squared shell easily because this possibility requires two massive stars. However, in this case, it is difficult to explain the non-detection of the two massive stars at any wavelength, from optical to centimeter. Furthermore, previous studies in this region report a bolometric luminosity for NGC\,6334A of $\sim$5000~$L_\odot$ that is lower than expected for a massive-binary system with an O8 member \citep[e.g.][]{Tige2017,palau2021}.

On the other hand, the runaway star scenario requires a star less massive than in the CWR scenario. According to \citet{Pereira2016}, a non-thermal bow shock could be produced by a star with a mass-loss rate of $\gtrsim 10^{-7}$~$M_\odot$\,yr$^{-1}$. Based on Fig. 1d of \citet{vink2001}, such a mass-loss rate could be produced by a main-sequence star with solar metalicity and effective temperature $\gtrsim 12000$~K, corresponding to a star of spectral type earlier than B8 \citep[e.g.,][]{fitzpatrick07}. This result is consistent with the low luminosity reported by \citet[]{Tige2017} and \citet{palau2021} in this region. The proper motion of \tcrs\ is consistent with the velocities in the plane of the sky of previous observations of massive-runaway stars, such as AE Aurigae, which has a velocity of 150 km s$^{-1}$ or HD~104565, with a velocity of 193.7 $\pm$ 25.2 km s$^{-1}$ \citep[e.g.][]{Lin2023,carreterocastillo2023}. In this scenario the non-detection of the optical/infrared counterpart of the involved massive star is a less severe problem because only one star is required, and of less mass. However, the centimeter emission of the shell, of around 9 Jy at 23~GHz \citep[e.g.,][]{Li2020} is not easily explained in this case. In the scenario of the runaway star, the shell would be produced by a dynamic or explosive event driven by an encounter of multiple stars, which generated an important amount of energy, similar to the Orion-KL case \citep{zapata2009}. In this scenario, the shell could emit both thermal and non-thermal emission at the same time. In Appendix \ref{App:coolingshell}, a preliminary exploration of the dynamical time of a swept-up shell produced by a dynamic event is presented, assuming initial parameters consistent with the \tcrs\ properties. The dynamical time is found to be around 300--400 yr, comparable to the cooling time of the shell, of $\sim200$~yr, consistent with the fact that the shell is still observed at centimeter wavelengths. In addition, the estimated dynamical time is comparable to the dynamical time of the explosive event in Orion-KL, of $\sim500$~yr \citep[e.g.,][]{zapata2009,rodriguez2017}. Thus, this scenario cannot be discarded and deserves to be studied in more detail in the future to assess more robustly its feasibility.

\section{Conclusions}

The high angular resolution observations presented in this work
have been used to study \tcrs, the compact radio source located in the 
center of the C-HII region \ngca. The source is resolved, and its morphology and spectral index of $-0.68\pm0.17$ suggest that 
it could be a CWR or a bow shock produced by a runaway star.

In the CWR scenario, \ngca\ would be ionized by at least two young massive stars that, however, have not been detected at any wavelength. The two young stars have to be of spectral type O or early B, massive enough so that the interaction of their winds can accelerate particles to  relativistic speeds. Classically, CWRs with non-thermal radio emission have mostly 
been studied in binary systems where at least one of the components 
is a WR star. WR stars are evolved massive O stars that have left
the main sequence phase and are in the latest phases of the star 
evolution. Cases of CWRs whose involved stars are in the main sequence phase are even less common.
In principle, young massive binary systems inside compact HII regions could also produce CWRs. However, as the non-thermal emission can be easily absorbed by the thermal radio emission from the surrounding HII region and the fainter emission from the CWR could be dominated by that of the HII region, they can be hard to detect. The case of \tcrs\ is interesting since the shell-like structure of \ngca\ could open a window to observe and study this CWR at early stages of massive star formation.  
If this scenario is confirmed, this would be the first time that a CWR is detected in such an early phase of star formation. 


Alternatively, the source could be the bow shock of a runaway star, probably of B-type, given the fact that the source is not detected in the optical/infrared. This is consistent with the preliminary analysis of proper motions presented in this work, which seems to indicate that \tcrs\ is moving towards the south at around 100 km\,s$^{-1}$. These results need to be confirmed with future observations that would allow to cover a wider time span, and would be specifically designed to measure the proper motions with the same phase calibrator in all epochs. If this scenario was confirmed, \tcrs\ would result from a dynamical interaction of several stars, that might have caused a shock wave that heated and ionized the surrounding medium. This dynamical scenario has recently been found to be more common than previously thought \citep[e.g.,][]{carretero23}. However, it is not clear if such a scenario could explain the large amount of centimeter emission associated with the shell. This scenario was preliminarily explored in the Appendix of this work, providing reasonable cooling timescales comparable to the timescales of dynamical events identified in other massive star-forming regions. Further theoretical work would help discern the feasibility of the dynamical event and runaway scenario.

Overall, the present study of \tcrs\ revealed that this source is an excellent opportunity to explore and learn about important processes in the formation of massive stars, such as colliding winds or dynamical events, and constitutes an excellent laboratory for future studies.

\section*{Acknowledgements}

The authors are grateful to Jakob Van den Eijnden for very stimulating discussions, and to the anonymous referee for a critical reading of the manuscript that improved the quality of the paper.
V. Y. acknowledges financial support from CONAHCyT, M\'exico, and UNAM. V. Y. and A.P. acknowledge financial support from the UNAM-PAPIIT IG100223 grant. 
A. P. acknowledges support from the Sistema Nacional de Investigadores of CONAHCyT, and from the CONAHCyT project number 86372 of the `Ciencia de Frontera 2019’ program, entitled `Citlalc\'oatl: A multiscale study at the new frontier of the formation and early evolution of stars and planetary systems’, M\'exico.
S.A.D. acknowledges the M2FINDERS project from the European Research
Council (ERC) under the European Union's Horizon 2020 research and innovation programme
(grant No 101018682).
The National Radio Astronomy Observatory is a facility of the National Science Foundation operated
under cooperative agreement by Associated Universities, Inc.
This scientific work uses data obtained from Inyarrimanha Ilgari Bundara / the Murchison Radio-astronomy Observatory. We acknowledge the Wajarri Yamaji People as the Traditional Owners and native title holders of the Observatory site. CSIRO’s ASKAP radio telescope is part of the Australia Telescope National Facility (https://ror.org/05qajvd42). Operation of ASKAP is funded by the Australian Government with support from the National Collaborative Research Infrastructure Strategy. ASKAP uses the resources of the Pawsey Supercomputing Research Centre. Establishment of ASKAP, Inyarrimanha Ilgari Bundara, the CSIRO Murchison Radio-astronomy Observatory and the Pawsey Supercomputing Research Centre are initiatives of the Australian Government, with support from the Government of Western Australia and the Science and Industry Endowment Fund. This paper includes archived data obtained through the CSIRO ASKAP Science Data Archive, CASDA (https://data.csiro.au).
This document was prepared using the collaborative tool Overleaf available at 
\url{https://www.overleaf.com/}. 



Additional to the software mentioned in the paper we have also used the following software: 
Astropy\footnote{\url{https://www.astropy.org/}} \citep{astropy2013}, NumPy\footnote{\url{https://www.numpy.org/}} \citep{numpy2011}, SciPy\footnote{\url{https://www.scipy.org/}} \citep{scipy2014}, Matplotlib\footnote{\url{https://matplotlib.org/}} \citep{matplotlib2007}, and APLpy \citep{aplpy2012}. 

\section*{Data Availability}

The data underlying this paper will be shared on request to the corresponding author.




\bibliographystyle{mnras}
\bibliography{mnras} 




\appendix

\section{The catalog of compact radio sources in the NGC 6334A region}\label{App:CRS}

Additional to \tcrs, other compact radio sources\footnote{Unresolved or slightly resolved sources.} are detected in the final maps. While not discussed in this paper, these sources are interesting on their own, as they may be a representative population of stars associated with \ngca. Here we describe the steps taken to obtain important 
parameters of these sources.

The source extraction was carried out using the BLOBCAT software \citep[see][]{hales2012}. This software extracts compact and extended sources from astronomical images. It applies bias correction of Gaussian and non-gaussian fits, obtaining the properties of the sources. It has been applied to extract the radio sources of many previous works \citep[e.g., ][]{dzib2018,medina2018,medina2019,yanza2022}. For the extraction, BLOBCAT needs a noise background map, which is obtained using the Graphical Astronomy and Imaging Analysis Tool \citep[GAIA, based on SExtractor;][]{bertin1996,holwerda2005}. 
Since the images have extended emission and high noise in some areas, we used a starting source detection limit of 5$\sigma$.

To ensure the reliability of the detection from BLOBCAT, we first considered sources with a signal-to-noise ratio above 7 as highly reliable. Then, we did a position cross-match of sources detected in the different bands.
Sources are considered as reliable if they have a detection level of $7\sigma$ in at least one of the maps.   The results are shown in Table \ref{tab:crs}. 

\begin{figure*}
 \centering
    \includegraphics[height=0.7\textwidth]{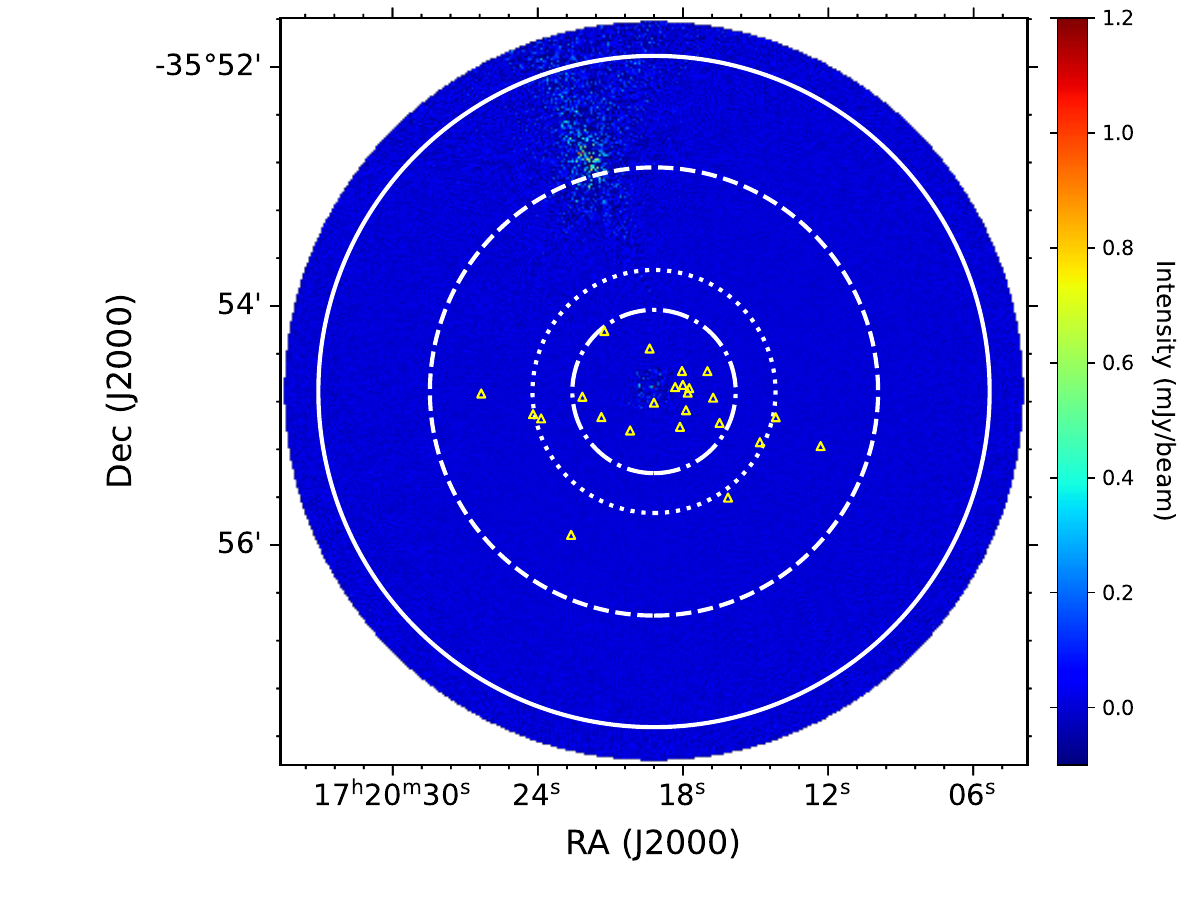}
     \caption{Compact radio sources (shown as yellow triangles) around \ngca. {\it Background:} X-band image combining the three observed epochs. Concentric circles represent the different primary beam sizes of observed bands: dashed-dot line for Ka-band (33.0 GHz), dotted line for K band (22.0 GHz), dashed line for the upper side frequency of X-band (12.0 GHz), and solid line for the lower side frequency of X-band (8.0 GHz). The bright source in the northern part of the image is the quasar NGC 6334B.}
 \label{fig:Full}
\end{figure*}

Figure \ref{fig:Full} shows the position of the 24 compact radio sources detected by BLOBCAT in the three bands (those sources shown in Table \ref{tab:crs}). The background image corresponds to the X band map. Most of the sources are close to the phase center. 

The expected number of extragalactic sources were calculated following the same analysis made in \citet{yanza2022}. In this work, we used the formulation derived by \citet[their equation A11]{anglada1998}. The maximum expected number of extragalactic sources $\langle N \rangle$ was found using a central frequency of 10~GHz (X band has the largest field of view), $S_{0}  = 0.05$~mJy (the maximum upper limit of an undetected source in the X-band) and $\theta= 6\rlap{.}'2$. Using these parameters we obtain  $\langle N \rangle=1.5\pm1.2$, then we expect up to three extragalactic background sources in the imaged region. So far, only NGC~6334B is confirmed to be an extragalactic source, then only two of the 24 compact radio sources may be background sources. 
Thus, most of the detected sources are expected to be related to \ngca. 


 \begin{table*}
\small
\setlength{\tabcolsep}{3.5pt}
  \caption{Compact radio sources detected in the direction of NGC 6334A.}
  \label{tab:landscape}
  \begin{tabular}{ccccccccccc}
    \hline
    \hline
    R.A.\supb & Dec.\supb & \multicolumn{4}{c}{Peak Intensity (mJy beam$^{-1}$)} &\multicolumn{4}{c}{ Flux Density (mJy)}&  \\ \cmidrule(lr{.75em}){3-6} \cmidrule(lr{.75em}){7-10}
(J2000)&(J2000)&X$_{\rm all}$\supa &K  &K+Ka &Ka & X$_{\rm all}$\supa &K  &K+Ka &Ka & $\alpha$ \\
    \hline
17 20 12.315& -35 55 10.38& 0.07$\pm$0.01 & \nodata & \nodata & \nodata & 0.07$\pm$0.01 & \nodata & \nodata & \nodata &  \\
17 20 14.169& -35 54 55.92& 0.13$\pm$0.01 & $<$0.2$\pm$0.04 & $<$0.14$\pm$0.03 & \nodata & 0.11$\pm$0.01 & $<$0.2$\pm$0.04 & $<$0.14$\pm$0.03 & \nodata &  \\
17 20 14.824& -35 55 8.44& 0.19$\pm$0.01 & $<$0.18$\pm$0.04 & $<$0.13$\pm$0.03 & \nodata & 0.17$\pm$0.01 & $<$0.18$\pm$0.04 & $<$0.13$\pm$0.03 & \nodata &  \\
17 20 16.144& -35 55 36.22& $<$0.06$\pm$0.01 & 0.82$\pm$0.06 & $<$0.17$\pm$0.03 & \nodata & $<$0.06$\pm$0.01 & 0.84$\pm$0.06 & $<$0.17$\pm$0.03 & \nodata &  \\
17 20 16.490& -35 54 58.76& 0.32$\pm$0.02 & $<$0.13$\pm$0.03 & $<$0.07$\pm$0.01 & $<$0.08$\pm$0.02 & 0.28$\pm$0.02 & $<$0.13$\pm$0.03 & $<$0.07$\pm$0.01 & $<$0.08$\pm$0.02 &  \\
17 20 16.757& -35 54 46.08& 0.12$\pm$0.01 & $<$0.12$\pm$0.02 & $<$0.06$\pm$0.01 & $<$0.07$\pm$0.01 & 0.12$\pm$0.01 & $<$0.12$\pm$0.02 & $<$0.06$\pm$0.01 & $<$0.07$\pm$0.01 &  \\
17 20 16.994& -35 54 32.76& 0.14$\pm$0.02 & $<$0.11$\pm$0.02 & $<$0.06$\pm$0.01 & $<$0.07$\pm$0.01 & 0.18$\pm$0.02 & $<$0.11$\pm$0.02 & $<$0.06$\pm$0.01 & $<$0.07$\pm$0.01 &  \\
17 20 17.743& -35 54 41.32& $<$0.1$\pm$0.02 & $<$0.1$\pm$0.02 & 0.09$\pm$0.01 & 0.08$\pm$0.01 & $<$0.1$\pm$0.02 & $<$0.1$\pm$0.02 & 0.09$\pm$0.01 & 0.09$\pm$0.01 &  \\
17 20 17.807& -35 54 43.59& $<$0.1$\pm$0.02 & 0.11$\pm$0.02 & 0.15$\pm$0.01 & 0.15$\pm$0.02 & $<$0.1$\pm$0.02 & 0.12$\pm$0.02 & 0.17$\pm$0.01 & 0.21$\pm$0.02 & 1.4$\pm$0.6 \\
17 20 17.879& -35 54 52.39& 0.65$\pm$0.04 & 0.42$\pm$0.03 & 0.12$\pm$0.01 & $<$0.06$\pm$0.01 & 0.87$\pm$0.05 & 0.44$\pm$0.03 & 0.27$\pm$0.02 & $<$0.06$\pm$0.01 &  -0.9$\pm$ 0.3 \\
17 20 18.009& -35 54 39.65& $<$0.05$\pm$0.01 & $<$0.1$\pm$0.02 & 0.07$\pm$0.01 & 0.09$\pm$0.01 & $<$0.05$\pm$0.01 & $<$0.1$\pm$0.02 & 0.07$\pm$0.01 & 0.1$\pm$0.01 &  \\
17 20 18.044& -35 54 32.68& $<$0.1$\pm$0.02 & $<$0.1$\pm$0.02 & 0.19$\pm$0.01 & 0.16$\pm$0.02 & $<$0.1$\pm$0.02 & $<$0.1$\pm$0.02 & 0.21$\pm$0.01 & 0.16$\pm$0.02 &  \\
17 20 18.126& -35 55 00.70& $<$0.05$\pm$0.01 & $<$0.11$\pm$0.02 & 0.07$\pm$0.01 & 0.1$\pm$0.02 & $<$0.05$\pm$0.01 & $<$0.11$\pm$0.02 & 0.13$\pm$0.01 & 0.13$\pm$0.02 &  \\
17 20 18.326& -35 54 40.76& $<$0.1$\pm$0.02 & 0.15$\pm$0.02 & 0.11$\pm$0.01 & $<$0.05$\pm$0.01 & $<$0.1$\pm$0.02 & 0.15$\pm$0.02 & 0.12$\pm$0.01 & $<$0.05$\pm$0.01 &  \\
17 20 19.202& -35 54 48.61& 0.7$\pm$0.07 & 0.28$\pm$0.03 & 0.2$\pm$0.02 & 0.13$\pm$0.02 & 1.08$\pm$0.08 & 0.78$\pm$0.05 & 0.7$\pm$0.04 & 0.45$\pm$0.03 & -0.6$\pm$0.2 \\
17 20 19.380& -35 54 21.40& $<$0.1$\pm$0.02 & $<$0.11$\pm$0.02 & 0.09$\pm$0.01 & 0.07$\pm$0.02 & $<$0.1$\pm$0.02 & $<$0.11$\pm$0.02 & 0.12$\pm$0.01 & 0.13$\pm$0.02 &  \\
17 20 20.185& -35 55 02.55& 1.22$\pm$0.07 & 0.8$\pm$0.05 & 0.8$\pm$0.04 & 0.61$\pm$0.04 & 1.83$\pm$0.09 & 1.1$\pm$0.06 & 1.04$\pm$0.05 & 0.7$\pm$0.04 & -0.7$\pm$0.1 \\
17 20 21.252& -35 54 12.60& $<$0.11$\pm$0.02 & 0.27$\pm$0.03 & $<$0.08$\pm$0.02 & $<$0.1$\pm$0.02 & $<$0.11$\pm$0.02 & 0.21$\pm$0.03 & $<$0.08$\pm$0.02 & $<$0.1$\pm$0.02 &  \\
17 20 21.373& -35 54 55.80& 0.1$\pm$0.01 & $<$0.12$\pm$0.02 & $<$0.07$\pm$0.01 & $<$0.08$\pm$0.02 & 0.1$\pm$0.01 & $<$0.12$\pm$0.02 & $<$0.07$\pm$0.01 & $<$0.08$\pm$0.02 &  \\
17 20 22.158& -35 54 45.60& 0.2$\pm$0.01 & 0.26$\pm$0.03 & 0.1$\pm$0.01 & \nodata & 0.18$\pm$0.01 & 0.23$\pm$0.03 & 0.1$\pm$0.01 & \nodata &  0.3$\pm$0.4\\
17 20 22.625& -35 55 54.96& 0.08$\pm$0.01 & $<$0.39$\pm$0.08 & \nodata & \nodata & 0.07$\pm$0.01 & $<$0.39$\pm$0.08 & \nodata & \nodata &  \\
17 20 23.865& -35 54 56.49& 0.23$\pm$0.02 & 0.25$\pm$0.04 & 0.27$\pm$0.03 & $<$0.23$\pm$0.05 & 0.25$\pm$0.02 & 0.23$\pm$0.04 & 0.35$\pm$0.03 & $<$0.23$\pm$0.05 &  -0.1$\pm$0.6\\
17 20 24.198& -35 54 54.35& 0.16$\pm$0.01 & $<$0.22$\pm$0.04 & $<$0.17$\pm$0.03 & \nodata & 0.3$\pm$0.02 & $<$0.22$\pm$0.04 & $<$0.17$\pm$0.03 & \nodata &  \\
17 20 26.338& -35 54 43.99& 0.11$\pm$0.02 & \nodata & \nodata & \nodata & 0.15$\pm$0.02 & \nodata & \nodata & \nodata &  \\
    \hline
  \end{tabular}
  \begin{list}{}{}
\item[\supa] Values obtained from the image combining the three epochs where the X-band was observed.
\item[\supb] The mean position error is $\pm$0.01''. Position units are $^{h}\,^{m}\,^{s}$ for right ascension, and $^{\circ}\,'\,''$ for declination.
\end{list}
\label{tab:crs}
 \end{table*}

\section{Cooling shell after a dynamical event}\label{App:coolingshell}

This Appendix considers the scenario of a single energy injection caused by a dynamic interaction between multiple stars.
The observed shell of Fig.~\ref{fig:square} could be explained through a shockwave produced by the isotropic energy release after the dynamical event. For this, several parameters need to be estimated from the observed shell. These are the turnover frequency, the molecular density of the medium, the electron density and the shell width. 

\subsection{Spectral energy distribution}

\begin{figure}
 \centering
    \includegraphics[height=0.35\textwidth]{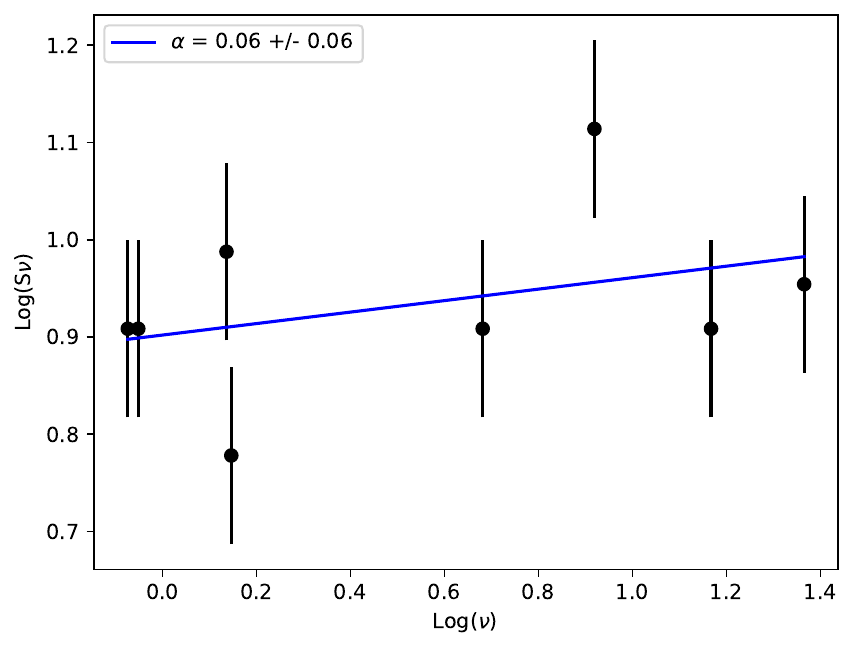}
     \caption{Spectral Energy Distribution in NGC~6334A from archive data of SUMSS, NVSS and VLA, with the fluxes taken within an aperture of $\sim60''$.}
 \label{fig:SED}
\end{figure}

For a single spherical uniform HII region, the spectral energy distribution (SED) is expected to have a spectral index of +2 when it is optically thick and $-0.1$ when it is optically thin. For more complex regions, the index could be in between these two values \citep[e.g.,][]{sanchez2011phd}. In Fig.~\ref{fig:SED} the SED of NGC~6334A is presented, covering the frequency range from 0.8 to 23 GHz, which is based on interferometric observations with low angular resolution, essentially from \citet{depree1995} using the VLA in D configuration, the NRAO VLA Sky Survey (NVSS), the Rapid ASKAP Continuum Survey (RACS) and the Sydney University Molonglo Sky Survey (SUMSS). Table \ref{tab:seddata} lists the fluxes and references that were used. This figure shows an almost flat index at the different frequencies, indicating that the emission is in the optically thin regime. Therefore, the possible turnover frequency for the shell should be around 1\,GHz or lower. 

\begin{table*}
    \centering
    \caption{Archive data of low angular resolution observations in NGC~6334A.}
    \begin{tabular}{c c c c l}
    \hline \hline
         Survey or Telescope & Frequency & Flux Density\supa & Synthesized beam & Reference\supb \\
           & (GHz) & (Jy) &  & \\
         \hline
         SUMSS & 0.864 & 8.1 & $43''$ & \citet{bock1999}\\
         RACS & 0.887 & 8.1 & $25''$ & \citet{mccon2020,hale2021} \\
         RACS & 1.37 & 9.7 & $9\rlap{.}''7\times8\rlap{.}''5$,  $46\rlap{.}^{\circ}5$ & \citet{mccon2020,hale2021} \\
         NVSS & 1.4 & 6 & $45''$ &  \citet{condon1998} \\
         VLA & 4.8 & 8.1 & $33.3''$ & \citet{depree1995} \\
         VLA & 8.3 & 13 & $17\rlap{.}''3\times6\rlap{.}''0$,  $-7\rlap{.}^{\circ}0$ & \citet{depree1995} \\
         VLA & 14.7 & 8.1 & $4\rlap{.}''3\times2\rlap{.}''6$,  $16\rlap{.}^{\circ}0$  & \citet{depree1995} \\
         VLA & 23.2 & 9 & $4\rlap{.}''7\times1\rlap{.}''5$,  $39\rlap{.}^{\circ}8$ & \citet{Li2020} \\
         \hline
    \end{tabular}
  \begin{list}{}{}
\item[\supa] The flux density was measured within an aperture of $\sim 60''$.
\item[\supb] These references correspond to the papers describing the surveys or to the papers that have published these data. Apart from the fluxes from \citet{depree1995}, all other fluxes were measured in this work using archive images.
\end{list}
    \label{tab:seddata}
    
\end{table*}

\subsection{Density of molecular gas and electrons}

In this section, an estimate of the electron density of the shell is provided.
We use the equation below to estimate the emission measure EM \citep[e.g.,][]{sanchez2011phd}:

\begin{equation}
\left[\frac{\nu_t}{\mathrm{GHz}}\right]=0.627\left[\frac{EM}{10^6 \mathrm{~cm}^{-6} \mathrm{pc}}\right]^{0.48}\left[\frac{T_{\mathrm{e}}}{10^4 \mathrm{~K}}\right]^{-0.64} ,
\label{eq:turnoverfreq}
\end{equation}

\noindent where $\nu_t$ is the turnover frequency and $T_e$ is the kinetic temperature of electrons, for which we assume $\nu_t$\,=\,1~GHz and $T_e$\,=\,5000~K, respectively. Under these assumptions EM\,=\,1\,$\times10^{6}$~cm$^{-6}$ pc.

The density was obtained using equation \ref{eq:n_e} \citep[e.g.,][]{toolsradio}, 

\begin{equation}
    n_e = \sqrt{\frac{EM}{L}},
    \label{eq:n_e}
\end{equation}

\noindent where $n_e$ is the density of electrons in the region and $L$ is the size of the shell. Assuming spherical symmetry for the shell and the radius corresponding to the aperture where the flux densities of Table~\ref{tab:seddata} were obtained, of $30''$, the length would be $60''$ or 80400~au. 
With these assumptions, the electron density is 1640 cm$^{-3}$. 

Regarding the density of molecular gas, we first estimate the mass of the gas and dust where it is detected, i.e., in the surroundings of the shell.  
To do this, the following equation \ref{eq:mass} was used \citep[e.g.,][]{palau2013}:

\begin{equation}
    \left[\frac{M}{M_\odot}\right]=3.25 \times \frac{\mathrm{e}^{0.048 \nu / T_{\mathrm{d}}}-1}{\nu^3 \kappa_\nu} \times\left[\frac{S_\nu}{\mathrm{Jy}}\right]\left[\frac{D}{\mathrm{pc}}\right]^2 ,
    \label{eq:mass}
\end{equation}

\noindent where $M$ is the total mass of gas and dust, $\nu$ is the frequency, $T_d$ is the dust temperature, $S_\nu$ is the flux density, and $D$ the distance. For this case, {\it Herschel} at 350 $\mu$m was used. The flux density is $S_\nu$ = 2250 Jy measured in an aperture of similar size of the resolution of the NVSS data ($\sim 1'$). The dust temperature is $T_d$ = 40~K, taken from Table 3 of \citet{palau2021}. The dust opacity, $\kappa_\nu$ = 0.101~cm$^{2}$g$^{-1}$, was taken from \citet{ossenkopf1994} using a gas-to-dust mass radio of 100. The obtained mass is 372 $M_\odot$.

Finally, the density was estimated adopting a radius of 30'', which gives 2$\times 10^{5}$ cm$^{{-3}}$. This is of the order of the density obtained using the density power-law model at the corresponding radius by \citet{palau2021}, of 9 $\times 10^{4}$ cm$^{-3}$. We will use 1$\times 10^{5}$ cm$^{{-3}}$ as an average value for the next calculations. Using our estimates of the electron density and molecular gas density, the ionization fraction of the shell is 0.016. Once these parameters are defined, the cooling time can be estimated for this scenario.

\subsection{Swept-up shell model}

We aim at explaining the properties of the shell presented in Fig.~\ref{fig:square}. From the figure, it is known that the shell is  1000-3000~au wide and that it is  $x=7000$~au away from its center of symmetry. Using the Saha equation, the temperature that produces the calculated ionization fraction $x_i=0.016$ in a gas of $10^5$ cm$^{-3}$ is 

\begin{equation}
    \frac{x_i^2}{1-x_i}=\frac{1}{n}\left(\frac{2\pi m_e k_b T}{h^2}\right)^\frac{3}{2}\exp^{-\phi_H/{k_b T}},
\end{equation}

\noindent which can be solved to obtain $T=4000$ K.  

Now we can estimate the velocity of the shock that would produce this temperature by following eq.~7.12 of \citet{raga2021}

\begin{equation}
    T=1.13\times 10^5 K \left(\frac{u}{100\,\rm{km\,s}^{-1}} \right)^2,
\end{equation}

\noindent where $u$ is the shock velocity. Assuming the temperature inferred above, a lower limit for the shock velocity is $u_s=18$~km~s$^{-1}$.

Assuming ballistic motion, the shell has been moving during a time 

\begin{equation}
    t_b=\frac{x}{u_s},
\label{eq:tb}
\end{equation}

\noindent which gives $t_b=1870$ yr. Nevertheless, this ballistic assumption implies that the shell does not convert kinetic energy to thermal energy and, therefore, thermal energy should be lost through cooling processes, with a corresponding cooling time and length of $190$ yr, and $\sim 550$ au, respectively, taking into account that the shell has a density of $\sim 10^5$~cm$^{-3}$, a final temperature of 100~K\footnote{Although a final temperature of 100 K is assumed, the current temperature of the shell is higher because the gas is heated while it is expanding and interacting with new gas. However, since the shell progressively slows-down, the maximum temperature achieved in the shock is smaller as the shell expands.} and a velocity $u_s$\footnote{If a density one order of magnitude lower is used, of $\sim 10^4$~cm$^{-3}$, then the cooling time is 1500 yr and the cooling distance is 5500~au.} \citep[following the cooling function of][]{Richings2014}.  
Since this cooling time is much smaller than the dynamical time of 1870 yr, it does not seem likely that this mechanism can explain the observed centimeter emission. 

Therefore, a more realistic model that accounts for the mass swept up and heated by the passing shock is considered. We can consider that the shell has been sweeping the surrounding material as it moves forward. Then, at all times, the shell with mass $M$ and velocity $u$ should preserve its momentum 
\begin{equation}
    Mu=M_0u_0, 
    \label{eq:motion}
\end{equation}

\noindent which also limits the initial velocity $u_0$ at time $t=0$, when the mass was $M_0$. Then we can define the mass as a function of the initial mass as $M = \beta M_0$, and therefore, $u_0\sim \beta u$, where $\beta$ is a factor that accounts for the mass increase. At this point, it would be useful to propose a swept mass function, which can be obtained assuming a geometry for the expanding shell. If the shell swept through all the material it encountered, the mass added to it would be described by 

\begin{equation}
    M=M_0 \left(\frac{x}{x_0}\right)^\gamma,
    \label{eq:Mass}
\end{equation}
where $\gamma$ is a geometric parameter, that accounts for an spherical ($\gamma=3$), flat ($\gamma=2$) or linear material accretion ($\gamma=1$), which combined with equation \ref{eq:motion}, can be interpreted as 

\begin{equation}
    \left(\frac{x}{x_0}\right)^\gamma d\left(\frac{x}{x_0}\right)= \frac{u_0}{x_0} dt,
\end{equation}

\begin{equation}
   t=\frac{x_0}{u_0} \frac{\left(\frac{x}{x_0}\right)^{\gamma+1}-1}{\gamma+1},
\end{equation}
and, dividing by eq.~\ref{eq:tb} on both sides,

\begin{equation}
   \frac{t}{t_b}=\frac{x_0/x}{u_0/u} \frac{\left(\frac{x}{x_0}\right)^{\gamma+1}-1}{\gamma+1},
\end{equation}

\noindent or

\begin{equation}
   \frac{t}{t_b}= \frac{1-\beta^{-\frac{\gamma+1}{\gamma}}}{\gamma+1}.
\end{equation}

\begin{figure}
    \centering
    \includegraphics[width=1\linewidth]{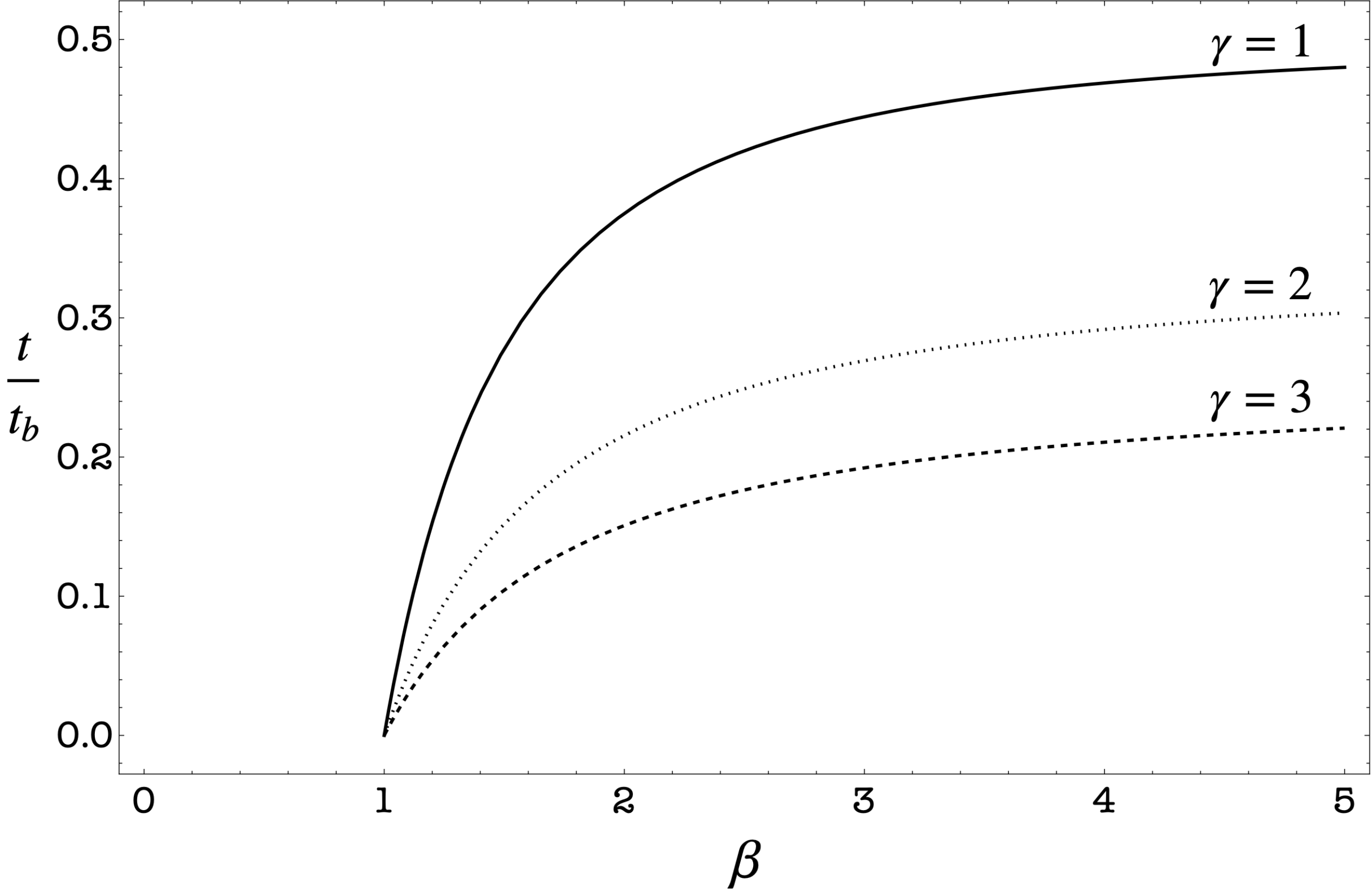}
    \caption{Life of the shockwake as function of the mass accreted for each expansion geometry. The dashed line corresponds to $\gamma=3$, the dotted line corresponds to $\gamma=2$, and the continuous line corresponds to $\gamma=1$.}
    \label{fig:tvsbeta}
\end{figure}

This last function approaches asymptotically to $(\gamma+1)^{-1}$ for a large $\beta$. In Figure \ref{fig:tvsbeta} we show the life of the shockwake as function of the mass accreted for each expansion geometry. 

Now, we can estimate the dynamical time for reasonable values of $\beta$, choosing an isotropic expansion, that is, $\gamma=3$. Taking into account that the dynamical event took place in a high-density medium, as observed in massive star-forming regions \citep[e.g.,][]{palau2021}, it would be expected that the swept-up mass and $\beta$ are large. For example, for $\beta=2$,
the dynamical time is $t=0.15 t_b\sim 280$ yr. Similarly, a very large $\beta$ corresponds to the asymptotic behavior, for which 
$t=0.2 t_b\sim 400$ yr. In both cases, this dynamical time is of the order of the cooling time calculated in previous paragraphs (of $\sim190$~yr), and therefore indicates that the swept-up shell scenario could be feasible for this region. 
In addition, the calculated cooling time is a lower limit because it assumed the ballistic case, while in the swept-up shell model, the expanding gas is continuously interacting with its surroundings and therefore continuously heating the gas as it expands, transforming kinetic energy into thermal energy. 
The cooling length of $550$ au would explain the thin shell walls observed, of $\sim1000$~au.
Besides, the dynamical time obtained from this swept-up shell model is comparable to the timescale estimated in regions where dynamical events have been identified, such as the explosive outflow of Orion-KL \citep[e.g.,][]{zapata2009, rodriguez2017}. In summary, the preliminary exploration performed here does not allow us to discard the dynamical event scenario for \tcrs.

\bsp	
\label{lastpage}
\end{document}